\documentclass[11pt]{revtex4}
\usepackage{microtype}
\usepackage{amsthm}
\usepackage{amsmath}
\usepackage{amsfonts}
\usepackage{MnSymbol}
\usepackage{bbold}
\usepackage[mathscr]{eucal}
\usepackage[textwidth=3cm,textsize=tiny,shadow]{todonotes}
\usepackage{natbib}
\usepackage{graphicx}
\usepackage{xcolor}
\usepackage{mathtools}
\usepackage{algorithm}
\usepackage{algpseudocode}
\begin{document}
\title{A novel approach to the spectral problem in the two photon Rabi model}
\author{Andrzej J.~Maciejewski} \email{maciejka@astro.ia.uz.zgora.pl}
\affiliation{Janusz Gil Institute of Astronomy, University of Zielona
  G\'ora, Licealna 9, PL-65--417 Zielona G\'ora, Poland.}%
\author{Tomasz Stachowiak} \email{stachowiak@cft.edu.pl}
\affiliation{%
  Center for Theoretical Physics, Polish Academy of Sciences, Al. Lotnikow 32/46, 02-668
  Warsaw, Poland }%
\affiliation{Department of Applied Mathematics and Physics, Graduate School of Informatics, Kyoto University, 606-8501 Kyoto, Japan} 

\begin{abstract}
  We explore the spectral problem of the two photon Rabi model from
  the point of view of complex differential equations in the Bargmann
  representation. The wave-functions are automatically entire but to
  ensure finite norm one has to effectively construct asymptotic
  expansions. This is achieved by means of the Mellin transformation
  and convergent factorial series, which allow direct computation of
  the spectral determinant.  By further analysing the differential
  equation satisfied by the Mellin transform, we obtain a new form of
  the spectral conditions -- in terms of holonomy matrices and contour
  integrals.
\end{abstract}

\maketitle

\section{Introduction}
Although the Bargmann-Fock representation is commonly applied in
quantum optics, its usage is often reduced to algebraic manipulation
of infinite matrices and Fock basis, while its analytical aspects
remain unexplored. In this article, we study the two photon Rabi
model, which is a particularly challenging example when formulated in
the language of complex differential equations. The main difficulty is
that instead of standard boundary conditions, we have constraints on
the asymptotic behaviour of wave-functions in all directions of the
complex plane.

The main goal is to give an effective method of calculating spectra,
which is directly derived from the properties of the Bargmann space of
entire functions.  The condition that the norm of a state be finite is
not straightforward to use in practical calculations.  Our main result
is to recast that requirement into constraints on the holonomy group
of the differential equation obtained by the Mellin transform of the
original system. This approach provides a novel theoretical framework,
and additionally has a simple numerical implementation.

Behind this mathematical description the idea is very simple.  The
physical condition that the norm of an eigenstate is finite requires
that solutions of differential equations with ``good'' asymptotics
glue nicely.  Application of the Mellin transformation gives rise a
new system of differential equations for which the physical condition
is now -- the system has a solution which glue local solutions with
``good exponents''. That is exactly the same form as the quantisation
problem formulated e.g. for the Rabi model \cite{Maciejewski:14::}. At
this point one can apply a variant of arbitrary methods used for
studying such models. In this paper we propose a seemingly new one
based on holonomy group of the system.

Stated another way, we wish to show how mathematical objects such as
the analytical continuation, Mellin transform and asymptotic series
can be effectively used to obtain the spectrum of physical problems
whose Schroedinger equation can be considered on the complex plane. We
will demonstrate all the relevant steps using an example from quantum
optics.

The Hamiltonian of the two photon Rabi model has the form
\begin{equation}
  H = \omega a^{\dag}a +\frac{\omega_0}{2}\sigma_z +
  2g \left[(a^{\dag})^2+(a)^2\right]
  \sigma_x,
  \label{H2pp}
\end{equation}
where $\sigma_x$, $\sigma_y$, and $\sigma_z$ are the Pauli matrices;
$a^{\dag}$ and $a$ are photon creation and annihilation operators.  It
is also known as the two-photon Jaynes-Cummings model investigated
initially in \cite{Gerry:88::} and with a view to population
inversion in \cite{Penna:16::}. For very recent studies of the two-photon Rabi model we refer the reader to \cite{JPA16,JPA17} where a detailed references for this subject can be found. 

This system and the Rabi model were also extensively investigated in
the PhD thesis of C.~Emary~\cite{Emary:01::}. In the subsequent paper
~\cite{Emary:02::} the authors show that the system is
quasi-exactly-solvable, i.e., a finite number of eigen-states is known
explicitly.  However, for generic values of the physical parameters
the spectrum is determined by a dedicated numerical methods based on
diagonalisation. As noticed in the above references without the
rotating wave approximation the system is not integrable. 

On the other hand, in \cite{Travenec:12::} the author claims that the
system is solvable and that the eigen-energies are zeros of a certain
expression called $G$ function. Yet another scheme of a ``simple'' spectra
calculation for the Rabi two-photon model was proposed in \cite{PRA86} and
according to the authors it gives a much simpler expression for the $G$ function.
However, as we have shown \cite{Maciejewski:15::}, the $G$ function used in 
\cite{Travenec:12::}
 is identically zero, so it cannot be reliably used to
obtain the spectrum.
Note that the Frobenius method, or its variants, utilised in these papers  can be applied to obtain a solution or $G$ function  for
{\it any} such system and hence says nothing about solvability.  

This motivates the second goal of this article, which is to use the
method introduced here to derive, as explicit as possible, a spectral
determinant whose zeros are the eigen-energies of the two photon Rabi
model. At the same time, we want the derivation to follow from the basic formulation of a quantum eigen-problem, without heuristics or conjectures.

The complete analysis consists of several crucial steps. First, the
physical system has to be represented in purely mathematical terms,
which in this case are: a system of ordinary differential equations on
a complex plane and initial or boundary conditions. As it turns out,
the boundary conditions here are strictly speaking the asymptotic
behaviour at infinity. This initial translation into mathematical
language is given in section \ref{sect:Formulation}.

The problem of asymptotic behaviour of the solutions will then be
treated by means of the Mellin transform in section
\ref{sect:Mellin}. As the growth of an entire function can be given in
terms of the coefficients of its series expansion, the natural
language here is that of recurrence relations and the Mellin transform
is very useful in solving them. This will lead to direct spectral
conditions in terms of sequences instead of complex functions.

Although ready for applications, the recurrence approach has some
drawbacks and the sections \ref{sect:Contour} and \ref{sect:Holonomy}
are devoted to overcoming them. We will try to reformulate the
spectral condition in analytical, rather than discrete language,
giving it as a contour integral and discussing how the analytical
continuation leads to the consideration of the holonomy group. Its
action is the most important part of our study.

Finally, we will bring the elements of the analysis together
formulating a complete criterion in section \ref{sect:Criterion} and
giving a possible implementation in section \ref{sect:Algorithm}.

\section{Formulation of the problem}\label{sect:Formulation}

In the Bargmann-Fock representation, see~\cite{Bargmann:61::}, the
wave function of a two level system $\psi=(\psi_1, \psi_2)$ is an
element of Hilbert space
$\mathscr{H}^2=\mathscr{H}\times \mathscr{H}$, where $\mathscr{H}$ is
the Hilbert space of entire functions of one variable
$z\in\mathbb{C}$. The elegant connection with the standard picture is
that the annihilation and creation operators $a$, and $a^{\dag}$
become $\partial_z$ and multiplication by $z$, respectively, for
clearly $[\partial_z,z]=1$. The scalar product in $\mathscr{H}$ is
given by
\[
  \langle f,g\rangle=
  \dfrac{1}{\pi}\int_\mathbb{C}\overline{f(z)}g(z)e^{-|z|^2}\mathrm{d}
  (\Re(z))\mathrm{d}(\Im(z)).
\]
It is worth mentioning that this space was also introduced,
independently of Bargmann, by J. Newman and H. S. Shapiro
\cite{Newman:64::,Newman:66::}.

The Hilbert space $\mathscr{H}$ has several peculiar properties. Let
us mention two of them:
\begin{enumerate}
\item $f(z)\in\mathscr{H}$ does not imply that $f'(z)\in\mathscr{H}$.
\item $f(z)\in\mathscr{H}$ does not imply that $zf(z)\in\mathscr{H}$.
\end{enumerate}
To understand these rather strange properties we have to recall some
definitions and facts from the theory of entire functions, see
\cite{Levin:96::,Boas:54::}. If $f(z)$ is an entire function, then to
characterise its growth, the following function is used:
\begin{equation}
  M_f(r) := \underset{|z|=r}{\mathrm{max}}|f(z)|.
\end{equation}
We omit the subscript $f$ later on, because the investigated function
is known from the context. If for an entire function $f(z)$ we have
\begin{equation}
  \lim_{r\rightarrow\infty}\sup\frac{\ln(\ln M(r))}{\ln r} = \varrho,\quad
  \mathrm{with}\quad 0\leq \varrho \leq \infty,
\end{equation}
then $\varrho$ is called the order (or growth order) of $f(z)$. If,
further, the function has positive order $\varrho<\infty$ and
satisfies
\begin{equation}
  \lim_{r\rightarrow\infty}\sup\frac{\ln M(r)}{r^\varrho} =\sigma,
\end{equation}
then we say that $f(z)$ is of order $\varrho$ and of type $\sigma$.

Assume that $f(z)$ belongs to $\mathscr{H}$, then one can prove the
following facts \cite{Bargmann:61::}:
\begin{enumerate}
\item $f(z)$ is of order $\varrho\leq 2$.
\item If $\varrho=2$, then $f(z)$ is of type $\sigma\leq\tfrac12$.
\end{enumerate}
If $\varrho=2$ and $\sigma=\tfrac12$, then the question whether
$f(z)\in\mathscr{H}$ requires a separate investigation. Exactly in the
mentioned case when $f(z)\in\mathscr{H}$ but $f'(z)\notin\mathscr{H}$
the function is of order $\varrho=2$ and type $\sigma=\tfrac12$. For
additional details see \cite{Vourdas:06::}.

The usefulness of this representation can immediately be seen with the
harmonic oscillator, which represents the radiation. The
time-independent Schr{\"o}dinger equation for energy $E$ is simply
$H\psi(z) = z\psi'(z)=E\psi(z)$ and one immediately recovers the
orthonormal eigenbasis as $\{z^n/\sqrt{n!}\}_{n\in\mathbb{N}}$. The
connection with the usual space of square-integrable functions of $q$
is given by the integral transformation
\begin{equation}
  \psi(z) = \frac{1}{\sqrt[4]{\pi}}
  \int_{-\infty}^{\infty}\exp\left[-(z^2+q^2)/2+\sqrt{2}qz\right]\phi(q)\mathrm{d}q, 
\end{equation}
and the kernel is one of the forms of the generating function for the
Hermite polynomials. Each $z^n$ thus corresponds to the appropriately
normalised wavefunction $e^{-q^2/2}H_n(q)$. In this basis, the
annihilation operator $a$ is just an infinite matrix with entries on
the superdiagonal, so the Hamiltonian can be constructed as tensor
products of such matrices with the sigma matrices, giving a simple
band structure. This allows for direct numerical
diagonalization. However, the open question that we wish to tackle is
how to determine the spectrum with as explicit exact formulas as
possible while avoiding heuristic reasoning.

Now we want to write down the Schr\"odinger equation for the two
photon Rabi model. First, we apply a unitary transformation given by
$U = (\sigma_x+\sigma_z)/\sqrt2$ to the Hamiltonian~\eqref{H2pp},
which gives
\begin{equation}
  \widetilde H=  U^\dag H U = \omega a^{\dag}a +\frac{\omega_0}{2}\sigma_x + 
  2g\left[(a^{\dag})^2+a^2\right]\sigma_z= 2g \left\{ 2x a^{\dag}a +\mu\sigma_x + 
    \left[(a^{\dag})^2+a^2\right]\sigma_z\right\},
\end{equation}
where we set $\omega=4xg$, and $\omega_0 =4\mu g$. We rewrite the
rescaled Hamiltonian in matrix form
\begin{equation}
  \begin{aligned}
    K:= \frac{1}{2g} \widetilde{H} &= 2x a^{\dag}a +\mu\sigma_x +
    \left[(a^{\dag})^2+a^2\right]\sigma_z \\
    &= \begin{bmatrix}
      2x a^{\dag}a    + \left[(a^{\dag})^2+a^2\right] & \mu \\
      \mu & 2x a^{\dag}a - \left[(a^{\dag})^2+a^2\right]
    \end{bmatrix}.
  \end{aligned}
\end{equation}

The operator $K$ can be decomposed as
\begin{equation}
  K = 2x A^{\dag}A + \mu\sigma_x - 2x\sin^2(\eta),
\end{equation}
where
\begin{equation}
  A = \cos(\eta) a + \sin(\eta)a^{\dag}\sigma_z,\quad
  \sin(2\eta) = \frac{1}{x}.
\end{equation}
For a normalized eigen-state $\psi=(\psi_1,\psi_2)$ we then have
\begin{equation}
  \langle\psi|K|\psi\rangle= 2x\|A\psi\|^2 
  +\mu\langle\psi|\sigma_x|\psi\rangle -2x\sin^2(\eta) = E,
\end{equation}
which gives the constraint
\begin{equation}
  E \geq 
  =-\mu-\tan(\eta).
  \label{e_bound}
\end{equation}

In the Bargmann representation the stationary Schr\"odinger equation
$K\psi = E\psi$, has the form of the following system of differential
equations
\begin{equation}
  \begin{aligned}
    \psi_1''(z) +2x z\psi_1'(z)+(z^2-E)\psi_1(z)+
    \mu\psi_2(z) &= 0,\\
    \psi_2''(z) -2x z\psi_2'(z)+(z^2+E)\psi_2(z)- \mu\psi_1(z) &= 0.
  \end{aligned}
  \label{eq:sys1}
\end{equation}
In some calculations it is more convenient not to take the above
system, but rather the corresponding fourth order equation, obtained
by elimination of $\psi_2$,
\begin{equation}
  \label{n=2}
  \psi_1^{(\text{iv})} + ((2-4x^2)z^2+4x)\psi_1''
  +4z(1+E x -x^2)\psi_1'+
  (2-E^2+\mu^2-4xz^2+z^4)\psi_1 = 0.
\end{equation}

All solutions of this equation are entire, so the other component
$\psi_2$ obtained from the first equation of \eqref{eq:sys1} is also
entire. This way, the first requirement of the Bargmann picture is
identically satisfied. What remains to be checked is the finiteness of
the norm.

Essentially all the subsequent work is devoted to the analysis of the
behaviour of the solutions at infinity and the structure of the
holonomy group of the relevant Mellin transform (introduced in the
next section).

There are four basis solutions, and they can be numbered according to
their parity, which is generated by the transformation $\tau$ of the
state
\begin{equation}
  \tau\psi(z) = \sigma_x \psi(\mathrm{i}z).
\end{equation}
The Hamiltonian commutes with $\tau$ and the associated symmetry group
is $\mathbb{Z}_4$ isomorphic to $\{1,\tau,\tau^2,\tau^3\}$. For a
solution with parity $s$ one has
\begin{equation}
  \begin{split}
    \psi_1(\mathrm{i} z) &= s \psi_2(z),\\
    \psi_2(\mathrm{i} z) &= s \psi_1(z),
  \end{split}
\end{equation}
where $s\in\{+1,-1,\mathrm{i},-\mathrm{i}\}$. Because
$\tau^2\psi(z) = \psi(-z)$, the distinction between even and odd
solutions represents the $\mathbb{Z}_2$ subgroup of the symmetry and
can be used to simplify the calculation.

In what follows we consider the even case, which includes parities
$s=\pm1$; calculations for the odd case $s=\pm\mathrm{i}$ are
completely analogous and the main stages are given in Appendix
\ref{odd_app}.

Let us define the function $f(\xi)$ such that $f(z^2) := \psi(z)$, for
which system\eqref{eq:sys1} becomes
\begin{equation}
  \begin{aligned}
    4\xi f_1''(\xi) + 2(2x\xi+1)f_1'(\xi) + (\xi-E)f_1(\xi) +\mu f_2(\xi) &= 0\\
    4\xi f_2''(\xi) - 2(2x\xi-1)f_2'(\xi) + (\xi+E)f_2(\xi) -\mu
    f_1(\xi) &= 0.
  \end{aligned}
\end{equation}
This system has a regular singular point at $\xi=0$ but we are only
interested in its entire solutions, given by series convergent in the
whole complex plane
\begin{equation}
  f(\xi) = \sum_{n=0}^{\infty} \boldsymbol{a}_n \xi^n,
  \label{fc}
\end{equation}
whose vector coefficients
$\boldsymbol{a}_n= \left[a_n^{1},a_n^2\right]^T$ satisfy the matrix
recurrence relation
\begin{equation}
  2n(2n-1)\boldsymbol{a}_n = \begin{bmatrix}
    E -4x(n-1) & -\mu \\
    \mu & -E +4x(n-1)
  \end{bmatrix} \boldsymbol{a}_{n-1} - \boldsymbol{a}_{n-2}.
  \label{rec}
\end{equation}
The initial conditions of the two solutions in question are fixed,
save for a multiplicative constant, by the choice of parities. The
symmetry action is $\tau f(\xi) = \sigma_x f(-\xi)$ which leads to
$f_{\pm}(-\xi) =\pm\sigma_x f_{\pm}(\xi)$ or, in terms of the series
coefficients,
\begin{equation}
  \begin{aligned}
    \boldsymbol{a}_{0} &= [1,1]^T,\quad\text{for}\; s=+1,\\
    \boldsymbol{a}_{0} &= [1,-1]^T,\quad\text{for}\; s=-1.
  \end{aligned}
  \label{rec_init}
\end{equation}


The only problem left is if the state has finite norm, which comes
down to the aforementioned growth order and type. Given the series
expansion $\psi_1(z) = \sum_{k=0}^{\infty}c_k z^k$ of a solution of
\eqref{n=2}, these quantities can be calculated as the following
limits
\begin{equation}
  \varrho = \limsup_{k\rightarrow\infty} \frac{k\ln k}{\ln(1/|c_k|)},\qquad
  (\mathrm{e} \sigma\varrho)^{1/\varrho} = 
  \limsup_{k\rightarrow\infty} k^{1/\varrho}|c_k|^{1/k}.
  \label{ordtype}
\end{equation}
Note that because $\xi = z^2$, the coefficients satisfy
$c_{2k} = a_k^{1}$ for even functions, and $c_{2k+1} = a_k^{1}$ for
odd ones. The order of $\psi_1(z)$ is thus double that of $f_1(\xi)$,
while the types are the same.

All the possible orders and types of $\psi(z)$ can be checked quickly
by substituting a formal series of the form
\begin{equation}
  \psi_1(z)=\exp(\sigma z^{\varrho})z^{\rho}
  \left(1+\frac{A_1}{z}+\frac{A_2}{z^2}+\cdots\right)
\end{equation} 
into equation~\eqref{n=2}, which yields $\varrho=2$ and four possible
types
\begin{equation}
  \sigma\in\left\{ \frac12(x\pm\sqrt{x^2-1}),
    -\frac12(x\pm\sqrt{x^2-1}) \right\}. 
  \label{types}
\end{equation}
Since the order is 2, the requirement of finite norm constrains the
type to the disk $|\sigma|\leq1/2$. The case $\sigma=1/2$ is
exceptional and is considered in appendix \ref{sigma12}. The condition
$\sigma<1/2$ implies the first physical requirement $x>1$, and that
only two out of four types in \eqref{types} are suitable.

The key difficulty here is that it is a priori unknown which
particular solution $\psi(z)$ specified around $z=0$ has which growth
type $\sigma$. In terms of the recurrence relation, one has particular
solutions $\boldsymbol{a}_{n}$ with the initial conditions
\eqref{rec_init}, but their behaviour as $n\rightarrow\infty$ required
for calculating \eqref{ordtype} remains to be checked. In general, the
recurrence relation has 4 basis solutions $\boldsymbol{b}_n$ each with
a different behaviour at infinity. In our case they can be determined
to be of the form
\begin{equation}
  \boldsymbol{b}_n \sim n^{\alpha n+\rho}\beta^n \boldsymbol{A}(n^{-1/p}),
  \label{formal_rec}
\end{equation}
where $\boldsymbol{A}$ is an asymptotic formal power series with some
suitable integer $p$.  For a general method of finding the asymptotic
solutions of linear recurrence relations see \cite{Turrittin}.

Using methods of this reference, it can be shown that such a basis can
be taken to be {\it asymptotically simple} in the sense that its
members exhaust all possible behaviors at infinity, i.e., the sets
$\{\alpha,\rho,\beta, \boldsymbol{A}\}$. Consequently, each
$\boldsymbol{a}_{n}$ can be represented as a linear combination of
such asymptotically simple $\boldsymbol{b}_n$ and this decomposition
will provide information about asymptotics of $\boldsymbol{a}_{n}$,
$f(\xi)$ given by \eqref{fc} and hence also about $\psi(z)$.

The main tool to achieve this will be a modified Mellin transform
which gives {\it convergent} expressions for $\boldsymbol{a}_n$ in the
form of factorial series. Such expressions are both valid for finite
$n$ and have prescribed asymptotic behaviour, so they can be compared
with $\boldsymbol{a}_n$ obtained recursively from \eqref{rec_init}. As
the solutions of a linear recursion relation form a vector space, the
problem will come down to checking dependence of finite-dimensional
vectors.

\section{The Mellin transform}\label{sect:Mellin}

Following Okubo \cite{Okubo:63::}, to determine solutions of the above
difference equation with prescribed asymptotic behaviour, we will use
the integral representation
\begin{equation}
  \boldsymbol{b}_n = \mathcal{M}[\boldsymbol{v}]_n := \frac{1}{\Gamma(1+n/\varrho)}\int_{C}u^{n}
  \boldsymbol{v}(u)\mathrm{d}u,
  \label{Mellin}
\end{equation}
where, $\boldsymbol{v}=[v_1(u),v_2(u)]^T$, and the contour $C$ will be
chosen such that the integrand resumes its initial value after $u$ has
described $C$. When compared to \cite{Okubo:63::}, the index $n$ is
shifted by 1 and we have modified the argument of the $\Gamma$
function to reflect the behaviour of coefficients of an entire
function of order $\varrho$. This can be easily seen from the
asymptotics
\begin{equation}
  \frac{1}{\Gamma(n/\varrho)} \sim \sqrt{\frac{n}{2\pi \varrho}}\left(\frac{\mathrm{e}\varrho}{n}\right)^{n/\varrho},
\end{equation}
while the formulae \eqref{ordtype} gives a simple example of
coefficients of an entire function of order $\varrho$:
\begin{equation}
  c_n = \left(\frac{\sigma \mathrm{e}\varrho}{n}\right)^{n/\varrho}.
\end{equation}
In other words, the $\Gamma$ factor ensures the order of $\varrho$,
whereas the $\sigma^n$ factor, which specifies the type, will have to
be recovered from the integral, by using a suitable
$\boldsymbol{v}(u)$.

In our case, the order of $\psi(z)$ is 2, so we should use
$\Gamma(n/2)$ to analyse \eqref{eq:sys1} or \eqref{n=2} directly, but
thanks to the separation of even and odd solutions, we can instead
deal with $f(\xi)$, which has order 1.

Thus, substituting the Mellin transform with $\varrho=1$ into relation
\eqref{rec}, we can transform it into a differential equation with
integration by parts of the form
\begin{equation}
  \int_C n^l u^{n-1} \boldsymbol{v}(u)\mathrm{d}u = [n^{l-1}u^n \boldsymbol{v}(u)]_C 
  - \int_C n^{l-1} u^n \boldsymbol{v}'(u)\mathrm{d}u.
  \label{intBpart}
\end{equation}
This allows to factor the integrand so that in the end the difference
equation becomes
\begin{equation}
  \begin{aligned}
    \frac{1}{\Gamma(n)}\int_C u^{n-1}\left((4u^2+4x\sigma_z
      u+1)\boldsymbol{v}'
      +(6u+(E+4x)\sigma_z)-\mathrm{i}\mu\sigma_y)\boldsymbol{v}\right) \mathrm{d}u&\\
    -\frac{1}{\Gamma(n)}\left[u^{n-1}(4u^2+4x\sigma_z
      u+1)\boldsymbol{v}\right]_C &= 0,
  \end{aligned}
  \label{inteq}
\end{equation}
so the system to solve for $\boldsymbol{v}$ is
\begin{equation}
  \frac{\mathrm{d}\boldsymbol{v}}{\mathrm{d}u} =  M (u) \boldsymbol{v}, \quad
  \setlength{\arraycolsep}{5pt}
  \setlength{\delimitershortfall}{-3pt}
  M (u) := -\begin{bmatrix}
    \dfrac{6u+4x+E}{4u^2+4xu+1} & \dfrac{-\mu}{4u^2+4xu+1} \\[3ex]
    \dfrac{\mu}{4u^2-4xu+1} & \dfrac{6u-4x-E}{4u^2-4xu+1} 
  \end{bmatrix}.
  \label{Mellin_sys}
\end{equation}
This system has five regular singular points $u_0$
\begin{equation}
  u_0\in\Omega=\left\{ \pm\frac{\kappa}{2}, \pm\frac{1}{2\kappa}, \infty\right\}, \quad\text{where}\quad x=\frac{\kappa}{2}+\frac{1}{2\kappa},
\end{equation}
and by the restrictions on $x$ we choose $0<\kappa<1$.  The
characteristic exponents at these points are the following
\begin{equation}
  \begin{aligned}
    \{0,-\chi\}, & \quad \text{for}\quad  u_0=\pm\frac{\kappa}{2}\\
    \left\{0,\chi-\frac32\right\}, &\quad \text{for}\quad    u_0=\pm\frac{1}{2\kappa}:\\
    \left\{-\frac32,-\frac32\right\}, &\quad \text{for}\quad u_0 =
    \infty,
  \end{aligned}
\end{equation}
where we introduced a natural spectral parameter
\begin{equation}
  \chi = \frac{\kappa(E+\kappa)}{2(1-\kappa^2)}+1.
\end{equation}
It is not a coincidence that the positions of these points are
precisely the growth types of the entire functions $f(\xi)$ and
$\psi(z)$, as will soon become apparent.  We also note that there is a
bound by \eqref{e_bound},
$\chi\geq -\frac{\mu\kappa}{2(1-\kappa^2)}+1$, because
$\kappa=\tan(\eta)$.

Now the question is how to distinguish local solutions appropriate for
the Mellin formula. On the one hand, the integral cannot vanish
identically, so the solution cannot be single valued if the contour is
a loop. On the other, the integrand must resume the same value on both
ends of $C$. Depending on $\chi$ then, we are lead to several
possibilities.

If $\chi$ is not a negative integer, then one local solution around
$u=\pm\kappa/2$ always has some kind of singularity: either there is a
branch point, a pole or a logarithmic term.  This allows for choosing
the contour $C$ simply as a loop starting at $u=0$ encircling a
specific singular point $\pm\kappa/2$ in the positive direction and
going back to zero. Choosing $\boldsymbol{v}(u)$ to be a solution
which is multivalued in a disk centred at $u_0$ will ensure that the
integral \eqref{Mellin} does not vanish identically, while the
boundary term in integration by parts will vanish at both ends,
i.e. $u=0$, provided that $u$ enters with positive power. The lowest
power boundary term in \eqref{inteq} is $u^{n-1}\boldsymbol{v}$;
consequently, such contour will work for at least $n\geq 2$.

If $\chi\in\mathbb{Z}_-$, the contour can be a line from the origin to
the singular point because the solution with the higher exponent has a
zero of order at least 1 at $u=\pm\kappa/2$, and both boundary terms
vanish.

In a neighbourhood of each singular point $u_0$ there exists a
solution with exponent $\nu$ given by the convergent series
\begin{equation}
  \boldsymbol{v}(u) = (u-u_0)^{\nu}\sum_{j=0}^{\infty} \boldsymbol{h}_j(u-u_0)^j.
  \label{serg}
\end{equation}
If the contour $C$ lies entirely in this neighbourhood, and the series
converges uniformly, the summation and integration can be exchanged,
and the Mellin transform yields the factorial series representation
\begin{equation}
  \begin{split}
    \mathcal{M}[\boldsymbol{v}]_n &= \frac{1}{\Gamma(n+1)}\int_C u^{n}
    \sum_{j=0}^{\infty}\boldsymbol{h}_j(u-u_0)^{j+\nu}\mathrm{d}u \\
    &= \frac{2\mathrm{i}\sin(\pi\nu)u_0^{n+1}}{\Gamma(n+1)}
    \sum_{j=0}^{\infty}\frac{(-u_0)^{\nu+j}\Gamma(n+1)\Gamma(1+j+\nu)}
    {\Gamma(2+j+n+\nu)}\boldsymbol{h}_j,
  \end{split}
  \label{fucktorial}
\end{equation}
which follows from the integral representation of the Beta function
\begin{equation}
  \frac{1}{2\mathrm{i}\sin(\pi\beta)}\int_C u^{\alpha-1}(u-1)^{\beta-1}\mathrm{d}u = 
  B(\alpha,\beta) = 
  \frac{\Gamma(\alpha)\Gamma(\beta)}{\Gamma(\alpha+\beta)}.
\end{equation}
If $C$ is a line, the $2\mathrm{i}\sin(\pi\nu)$ factor is absent.

The regular point used, $u_0$, determines the crucial asymptotic
behaviour and this is all one needs for practical purpose of finding
the ``good'' recurrence solutions $\boldsymbol{b}_n$ and gluing it with $\boldsymbol{a}_n$. This first method of obtaining the spectrum is described in appendix \ref{app_fac}.

Although $\boldsymbol{a}_n$ can be calculated explicitly for any $n$,
and the expressions for $\boldsymbol{b}_n$ are given by the factorial
series there are problems with direct computational
implementation. First, the complexity of the terms grows quickly with
$n$, and this makes the error estimates for $\boldsymbol{b}_n$
cumbersome. Second, because the singular points depend on the
parameters, the region $\kappa>1/\sqrt{2}$ requires separate
procedures of appendix \ref{was_tra}, for which the involved
expressions are even longer. Finally, we have to find a common zero of
4 functions ($3\times3$ minors) to locate the spectrum.

An alternative approach is based on the observation that
$\boldsymbol{b}_n$ are given by contour integrals, so they can be
calculated by numerical integration of $\boldsymbol{v}$ instead of
numerical summation of series. As it turns out, we can do even more
than that by noticing how the $\mathbb{Z}_4$ symmetry is reflected in
the holonomy group. In the process of finding the appropriate solution
$\boldsymbol{v}$ for the integral, we discover a very concise form of
the spectral conditions.

\section{The spectral condition as contour integral}
\label{sect:Contour}

Given an appropriate solution $\boldsymbol{v}$, the contour integrals
provide successive values of $\boldsymbol{b}_n$ for all $n\geq 2$;
alternatively, the whole sequence can equally well be generated by
\eqref{rec} from just two consecutive elements $\boldsymbol{b}_{n_0}$
and $\boldsymbol{b}_{n_0+1}$ obtained from the contour integrals. The
asymptotic growth of $\boldsymbol{b}_n$ is guaranteed by the above,
but the representation fails at $\boldsymbol{b}_0$, and in particular
it is not always the case that $\boldsymbol{b}_n\equiv0$ for $n<0$
contradicting the initial conditions \eqref{rec_init}. However, only
when they hold can we say that we have found the coefficients of a
function $f(\xi)$ that is both entire and of proper asymptotic type.

In order to express the initial conditions in terms of the function
$\boldsymbol{v}$, we can integrate the whole system \eqref{Mellin_sys}
over the contour in question. To this end we rewrite it as
\begin{equation}
  (4u^2+1)\boldsymbol{v'}+4xu\sigma_z \boldsymbol{v'} = -6u \boldsymbol{v} -(4x+E)\sigma_z\boldsymbol{v}
  +\mathrm{i}\mu\sigma_y \boldsymbol{v},
\end{equation}
and use integration by parts to get
\begin{align}
  -\int_{C}(8 u +4x\sigma_z)\boldsymbol{v} \mathrm{d}u +\int_C\boldsymbol{v}'\mathrm{d}u &=
                                                                        -\int_{C}\left[6u+(4x+E)\sigma_z -\mathrm{i}\mu\sigma_y\right]\boldsymbol{v}\mathrm{d}u, \\
  2\int_C u\boldsymbol{v}\mathrm{d}u  &= \int_C (E\sigma_z -\mathrm{i}\mu\sigma_y)\boldsymbol{v}\mathrm{d}u
                                        +[\boldsymbol{v}]_C,
                                        \label{int_sys}
\end{align}
where
$[\boldsymbol{v}]_C:=\boldsymbol{v}(\gamma(1))-\boldsymbol{v}(\gamma(0))$,
and $[0,1]\ni t\mapsto\gamma(t)$ is a parametrisation of the contour
$C$.  As we are in fact dealing with several singular points, we have
several pairs of contours and solutions
$\{C,\boldsymbol{v}\}_{u_0\in\Omega}$, in this particular case only
two points $u_0=\pm\kappa/2$ matter, so we will simply index the pairs
with $+$ and $-$. As explained in the previous section, the choice at
each point depends on the characteristic exponent, which also means
that each $\boldsymbol{v}$ is determined up to a multiplicative
constant.

A general solution $\boldsymbol{b}_n$ of the recurrence \eqref{rec}
can be any linear combination of particular solutions, each of which
corresponds to certain $\{C,\boldsymbol{v}\}_{u_0}$. Its asymptotic
type is determined by $u_0$ and to ensure that such $\boldsymbol{b}_n$
defines an entire function, it must coincide with $\boldsymbol{a}_n$
so the linear dependence
\begin{equation}
  \boldsymbol{a}_n = \alpha_+\boldsymbol{b}_n^++\alpha_-\boldsymbol{b}_n^-
\end{equation}
must hold. If so, then by the definition of $\boldsymbol{b}_n^{\pm}$
\begin{equation}
  \boldsymbol{a}_n = \frac{1}{n!}\int_{C_+}u^n\alpha_+\widetilde{\boldsymbol{v}}_+\mathrm{d}u+
  \frac{1}{n!}\int_{C_-}u^n\alpha_-\widetilde{\boldsymbol{v}}_-\mathrm{d}u,
  \label{abyMellin}
\end{equation}
and thanks to the aforementioned freedom of rescaling, we will write
$\boldsymbol{v}_{\pm} = \alpha_{\pm}\widetilde{\boldsymbol{v}}_{\pm}$.
For $n=1$ recurrence relation \eqref{rec} reads
$2\boldsymbol{a}_1 =
(E\sigma_z-\mathrm{i}\mu\sigma_y)\boldsymbol{a}_0$, which, by
substituting $\boldsymbol{a}_0$ and $\boldsymbol{a}_1$ as expressed in
\eqref{abyMellin}, becomes
\begin{equation}
  2\sum_i\int\limits_{C_i} u\boldsymbol{v}_i\mathrm{d}u = 
  (E\sigma_z-\mathrm{i}\mu\sigma_y)\sum_i\int\limits_{C_i}\boldsymbol{v}_i\mathrm{d}u. 
\end{equation}
Combining this condition with the integrated system \eqref{int_sys}
yields
\begin{equation}
  \sum_i\int\limits_{C_i}(E\sigma_z-\mathrm{i}\mu\sigma_y)\boldsymbol{v}_i\mathrm{d}u
  +\sum_i[\boldsymbol{v}]_{C_i} =
  \sum_i\int\limits_{C_i}(E\sigma_z-\mathrm{i}\mu\sigma_y)\boldsymbol{v}_i\mathrm{d}u ,
\end{equation}
or
\begin{equation}
  \boxed{\sum_i [\boldsymbol{v}_i]_{C_i} = 0.}
  \label{cond1}
\end{equation}
In the generic case, when the contour is a loop, this can be written
in terms of the holonomies
\begin{equation}
  \sum_i(F_i -\mathbb{1})\boldsymbol{v}_i(0) = 0,
\end{equation}
where, by definition, the holonomy matrix $F_i$ is the value of the
fundamental matrix $ V (u)$ of the system \eqref{Mellin_sys}
analytically continued over a closed loop $C_i$, starting with the
initial condition $ V (0)=\mathbb{1}$.

At this point we have replaced the need of constructing local series
representations of $\boldsymbol{v}$ around each singular point with
just obtaining the monodromy matrices for the system
\eqref{Mellin_sys}. The monodromy naturally incorporates the
information about characteristic exponents, which was also necessary
before to chose the right series and the contour. Moreover, we no
longer need to calculate several consecutive elements
$\boldsymbol{b}_n$ or even integrals
$\int u^n\boldsymbol{v}\mathrm{d}u$, because to obtain $m_i$, and
hence condition \eqref{cond1}, only one integration over $C_i$ is
sufficient.

\section{The impact of symmetry on the Holonomy}
\label{sect:Holonomy}

We notice first, that the matrix $ M (u)$ defining the right hand
sides of the the Mellin system \eqref{Mellin_sys} satisfies
\begin{equation}
  \label{symm}
  M (u)\sigma_x + \sigma_x  M (-u)={0}.
\end{equation}
Thus, if $\boldsymbol{v}(u)$ is a solution, so is
$\tau_M(\boldsymbol{v}(u)):=\sigma_x\boldsymbol{v}(-u)$.  We show that
this $\mathbb{Z}_2$ symmetry holds for any fundamental matrix, and
that its columns can be chosen as the eigenvectors of this symmetry.

Let ${T}(u)= {V}(u)^{-1}\sigma_x V(-u)$, where $ V(u)$ is an arbitrary
fundamental matrix of the Mellin system \eqref{Mellin_sys}, then
${T}'(u)={0}$. Direct differentiation gives
\begin{equation}
  \begin{aligned}
    {T}'(u) &=  \left[ V(u)^{-1}\right]'\sigma_x  V(-u) +  V(u)^{-1}\sigma_x\left[ V(-u)\right]' \\
    &= - V(u)^{-1} M (u)\sigma_x V(-u) + V(u)^{-1}\sigma_x\left[- M (-u) V(-u)\right]=\\
    \phantom{0}&\rule{2.75cm}{0cm}- V(u)^{-1}\left[ M (u)\sigma_x +
      \sigma_x M (-u)\right] V(-u)=0.
  \end{aligned}
\end{equation}
Hence ${T}={T}(u)$ is a constant matrix.

Now, assume that $V(0)=\mathbb{1}$, then $T=\sigma_x$
and hence
\begin{equation}
  \label{sxz}
  \sigma_x V(-u) =  V(u) \sigma_x. 
\end{equation}


Next, we investigate the holonomy  matrices of the system.
Let $a\in \mathbb{C}$ be a non-singular point of the system and $V(u)$ its local fundamental matrix defined in a neighbourhood of $a$. We consider analytic continuation  of initial state $Y_0=V(a)$ of the system along a loop $C$ given parametrically by the map 
\[
[0,1] \ni t \longmapsto \gamma(t)\in \mathbb{C},
\]
with $\gamma(0)=\gamma(1)=a$. The result of this continuation is a matrix $Y_1=Y(1)$ given  by the solution of initial value problem
\begin{equation}
    \frac{\mathrm{d}}{\mathrm{d} t} Y(t) =\dot\gamma(t)M(\gamma(t)) Y(t), \qquad Y(0)=Y_0.
    \label{eq_Y}
\end{equation}
This is a change of independent variable such that $V(\gamma(t)) = Y(t)$, and the fact that $t=0$ and $t=1$ both correspond to $u=a$, while $Y_0\neq Y_1$ in general, reflects the fact that $V(u)$ is not necessarily single-valued.

The holonomy matrix $F_{\gamma}$ is then defined via the linear map
\begin{equation}
    \Delta_{\gamma}: Y_0\mapsto Y_1 = F_{\gamma} Y_0,
\end{equation}
and it does not depend on the particular initial condition chosen, but only on the homotopy class of $\gamma$. This can be seen by noticing that any fundamental matrix of \eqref{eq_Y} is $Y(t)A$, for some constant matrix $A$, so that $Y_0A\mapsto Y_1A = F_{\gamma}Y_0 A$.

For further purposes we consider two parametrised loops $C_+$ and $C_-$  encircling counterclockwise the singular point $u= \pm\kappa/2$, respectively. They have one common point $u=0$. Loop $C_+$, 
parametrised by $\gamma_+(t)$,  gives the holonomy matrix 
$F_+$, i.e., $\dot{Y}=\dot\gamma_+ M(\gamma_+(t))Y$ and $Y_1=F_+Y_0$.
Similarly, we have $\dot{Z} = \dot\gamma_-M(\gamma_-(t))Z$ and $Z_1=F_-Z_0$. In order to find a relation between $F_+$ and $F_-$ we  assume  the loop $C_-$  is obtained form
$C_+$ by the reflection through the origin. Then $\gamma_-(t)=-\gamma_+(t)$ is
a parametrisation of $C_-$, and by \eqref{symm}
\begin{equation}
    \frac{\mathrm{d}}{\mathrm{d} t} (\sigma_x Z) = \sigma_x \dot{Z} =
    -\dot\gamma_+\sigma_xM(-\gamma_+) Z =
    \dot\gamma_+ M(\gamma_+) \sigma_x Z,
\end{equation}
which means that $\sigma_x Z(t) = Y(t)A$, for some constant matrix $A$, and 
\begin{equation}
\Delta_{\gamma_-}: Z_0\mapsto Z_1 = \sigma_x^{-1}Y_1A = \sigma_x^{-1}F_+ Y_0 A 
= \sigma_x^{-1} F_+\sigma_x Z_0,
\end{equation}
so we have obtained the fundamental formula
 \begin{equation}
  \boxed{F_- = \sigma_x^{-1} F_+\sigma_x.}
  \label{cond2}
\end{equation}

We also note that there is a direct link with the monodromy group, which is another representation of how the solutions change under analytic continuation along the contours. The monodromy matrices depend on the choice of the fundamental matrix but if the standard initial condition $V(0)=\mathbb{1}$ is used, they are numerically identical to the respective holonomies, so that an analogous formula $M_- = \sigma_x^{-1}M_+\sigma_x$ holds.

\section{Criterion}
\label{sect:Criterion}

We now bring together all the above elements to formulate criteria for
determining the spectrum. Thanks to the symmetry of the holonomy,
only analysis around one singular point is necessary; while the
contour formulation allows us to work with initial conditions
$\boldsymbol{v}(0)$.

{\bf Criterion 1.} {\it If $\chi$ belongs to the spectrum, i.e, the
  function $\psi(z)$ is entire and normalizable, then there exists a
  common eigenvector $\boldsymbol{e}$ of $F_+$, $F_-$ and $\sigma_x$.}

The local holonomy group is thus seen to be solvable because the matrices are simultaneously triangularizable. The above is just a necessary condition, and we further have

{\bf Criterion 2.} {\it Depending on the value of the parameters, the
  sufficient conditions are
  \begin{enumerate}
  \item $\chi\notin\mathbb{Z}$: the eigenvector $\boldsymbol{e}$ has
    the eigenvalue $\exp(2\pi\mathrm{i}\chi)$.

  \item $\chi\in\mathbb{Z}$ and there are logarithmic solutions or,
    equivalently, the holonomy has a Jordan block, the necessary
    condition is also sufficient.

  \item $\chi\in\mathbb{Z}$ and there are no logarithms or,
    equivalently, the holonomies are $F_{\pm}=\mathbb{1}$: the
    solution $\boldsymbol{v}_+$ corresponding to the eigenvector
    $\boldsymbol{e}$ has the characteristic exponent $-\chi$ at the
    regular point $\kappa/2$.
  \end{enumerate}}

To demonstrate all the points in turn, we will consider the initial
conditions of solutions $\boldsymbol{v}(0)$, denoted by
$\boldsymbol{e}$, and the action of the holonomy.

If $\chi\notin\mathbb{Z}$, the eigenvalues of $F_+$ are
$\lambda=\exp(2\pi\mathrm{i}\chi)$ and 1. Take the eigenvector
$\boldsymbol{e}^+_{\lambda}$ and construct
$\boldsymbol{e}^-_{\lambda}=-\alpha\sigma_x\boldsymbol{e}^+_{\lambda}$,
$\alpha\in\mathbb{C}^*$, which must be an eigenvector of $F_-$ to the
eigenvalue $\lambda$ by \eqref{cond2}. If the spectral condition
\eqref{cond1} is satisfied, then
\begin{equation}
  (F_+-\mathbb{1})\boldsymbol{e}^+_{\lambda} + (F_--\mathbb{1})\boldsymbol{e}^-_{\lambda}
  = (\lambda-1)\boldsymbol{e}^+_{\lambda}-\alpha(\lambda-1)\sigma_x \boldsymbol{e}^+_{\lambda} =0,
\end{equation}
but because $\lambda\neq1$ for noninteger $\chi$, it follows that
$\boldsymbol{e}^+_{\lambda}=\alpha\sigma_x\boldsymbol{e}^+_{\lambda} =
-\boldsymbol{e}^-_{\lambda}$. Thus, there is a common eigenvector of
$F_+$, $F_-$. In addition, it must be an eigenvector of $\sigma_x$, so
one of $[1,\pm 1]^T$.

If $\chi\in\mathbb{Z}$ and $F_+$ has a Jordan block, there is a
solution whose initial solutions at zero satisfy
$F_+\boldsymbol{e}^+_0=\boldsymbol{e}^+_0$ and a logarithmic solution
which corresponds to the generalized eigenvector, i.e.,
$F_+\boldsymbol{e}^+_l = \boldsymbol{e}^+_l +
2\pi\mathrm{i}\boldsymbol{e}^+_0$. Taking
$\boldsymbol{e}^-_i = -\alpha\sigma_x \boldsymbol{e}^+_i$,
$i\in\{0,l\}$, gives
\begin{equation}
  F_-\boldsymbol{e}^-_l = -\alpha\sigma_x \boldsymbol{e}^+_l -2\pi\mathrm{i}\alpha\sigma_x \boldsymbol{e}^+_0=\boldsymbol{e}^-_l+2\pi\mathrm{i}\boldsymbol{e}^-_0. 
\end{equation}
The spectral condition is then
\begin{equation}
  (F_+ -\mathbb{1})\boldsymbol{e}^+_l+
  (F_- -\mathbb{1})\boldsymbol{e}^-_l =
  2\pi\mathrm{i}(\boldsymbol{e}^+_0+\boldsymbol{e}^-_0)=0, 
\end{equation}
so these eigenvectors must be proportional and, like before,
$\boldsymbol{e}^+_0 = \alpha\sigma_x\boldsymbol{e}^+_0
=-\boldsymbol{e}^-_0$ is the common eigenvector of the form
$[1,\pm 1]^T$.

If $\chi\in\mathbb{Z}$ and $F_+$ is diagonalizable, it must be the
identity matrix, so the necessary condition is trivial; additionally
$\chi=0$ is excluded as it always leads to logarithms. For the
sufficient condition we notice, that for $\chi\in\mathbb{Z}_+$ and no
logarithms, the solution $\boldsymbol{v}(u)$ has a pole at the regular
point but it is not multivalued. The Mellin integral is thus not identically zero, but the contour condition $[\boldsymbol{v}]_C$ is identically satisfied around each point independently. As stated in section \ref{sect:Mellin}, this leads to pairs of explicit solutions discovered by Emary and Bishop
\cite{Emary:02::}. When $\chi$ is non-positive, $\boldsymbol{v}(u)$ has a zero
at the regular point and the contour has to be the line from $0$ to
$\kappa/2$. The corresponding solution around $-\kappa/2$ is
$\boldsymbol{v}_-(u):=-\alpha\sigma_x\boldsymbol{v}_+(-u)$ and the
contour condition \eqref{cond2} is
\begin{equation}
  \boldsymbol{v}_+(\kappa/2)-\boldsymbol{v}_+(0)+\boldsymbol{v}_-(-\kappa/2)-\boldsymbol{v}_-(0)=
  -\boldsymbol{e}^+-\boldsymbol{e}^- =0,
\end{equation}
and by the symmetry \eqref{cond1}, we must once again have
$\boldsymbol{e}^-=-\alpha\sigma_x\boldsymbol{e}^+=-\boldsymbol{e}^+$,
so that an eigenvector of $\sigma_x$ must correspond to the solution
with the positive exponent $-\chi$, i.e., vanishing at $\kappa/2$.

This completes the proof, and we also note that in the last case the
matrix $F_+$ cannot be used to obtain the eigenvector
$\boldsymbol{e}$; but to check which solution vanishes at $\kappa/2$ 
one can make use of Cauchy's integral
\begin{equation}
  \boldsymbol{v}\left(\tfrac{\kappa}{2}\right) =
  \frac{1}{2\pi\mathrm{i}} \oint \frac{\boldsymbol{v}(u)}{u-\frac{\kappa}{2}}\mathrm{d}u,
  \label{Cauchy}
\end{equation}
which will be valid for the whole fundamental matrix, since both
solutions are analytic.

In each of the above cases, the fundamental quantity is the
determinant
\begin{equation}
  \det[\boldsymbol{v}(u),\sigma_x\boldsymbol{v}(-u)],
\end{equation}
taken at $u=0$, where $\boldsymbol{v}$ is just $\boldsymbol{e}$, so
that if $\chi$ belongs to the spectrum
\begin{equation}
  W :=\det[\boldsymbol{e},\sigma_x\boldsymbol{e}] = 0.
\end{equation}

This determinant arises in complete analogy with the Wronskian
introduced by the authors in \cite{Maciejewski:15::}. Although here we
are dealing with a determinant of numeric quantities, these are the
initial conditions of solutions, and the connection the Wronskian of
$\boldsymbol{v}(u)$ is
\begin{equation}
  \text{Wr}[\boldsymbol{v}_1,\boldsymbol{v}_2] = W
  \frac{(1-4\kappa^2u^2)^{\chi-3/2}}{(1-4u^2/\kappa^2)^{\chi}}.
\end{equation}


\section{Implementation}
\label{sect:Algorithm}

All the cases can now be gathered into a simple algorithm for
computing the spectral determinant. The whole goal of finding the
correct values of the energy comes down to verifying that the main
equation has, for a given set of parameters $(E,x,\mu)$, entire
normalisable solutions. We will work directly with the quantities
$(\chi,\kappa,\mu)$, because they are more natural, e.g., the explicit
Emary-Bishop solutions appear for (half)integer values of $\chi$, and
$\kappa$ lies between 0 and 1.

We recall that the fundamental matrix $ V (u)$ has the initial
condition $ V (0)=\mathbb{1}$, and for numerical integration,
the contour around $\kappa/2$ can be parametrised with the path
\begin{equation}
  \gamma_+(t) = \frac14 - \frac14\exp[2\pi\mathrm{i} t],\quad t\in[0,1].
\end{equation}
The value that $ V $ attains at 0, having described the
contour $C$, will be the holonomy matrix $F_+$. For the exceptional
last case, we use Cachy's formula \eqref{Cauchy} for the whole matrix
$ V $ to obtain $ V (\frac{\kappa}{2})$, and its null
eigenvector will be the desired eigenvector $\boldsymbol{e}$.

\begin{algorithm}
\caption{Spectral determinant $W(\chi,\kappa,\mu)$}
  \begin{algorithmic}
    \Require{$\chi$, $\kappa$, $\mu$} \State Integrate system
    \eqref{Mellin_sys} to obtain $F_+= V (\gamma_+(1))$.
    \If{$\chi\notin\mathbb{Z}$} \State Determine the eigenvector
    $\boldsymbol{e}$ of $F_+$ to the eigenvalue
    $\mathrm{e}^{2\pi\mathrm{i}\chi}$.
    \ElsIf{$\chi\in\mathbb{Z}\;\land\;F_+\neq\mathbb{1}$} \State Take
    the only eigenvector $\boldsymbol{e}$.  \ElsIf
    {$\chi\in\mathbb{Z}_+\;\land\; F_+=\mathbb{1}$} \State Two
    Emary-Bishop states exist: $\boldsymbol{e}$ can be either of
    $[1,1]$ and $[1,-1]$.
    \ElsIf{$\chi\in\mathbb{Z}_-\;\land\;F_+=\mathbb{1}$} \State
    Integrate the fundamental matrix according to Cauchy's formula
    \eqref{Cauchy}.  \State Solve
    $ V (\kappa/2)\boldsymbol{e} = 0$ for $\boldsymbol{e}$.
    \EndIf
    \State $W = \det[\boldsymbol{e},\sigma_x\boldsymbol{e}]$.
  \end{algorithmic}
\label{algo}
\end{algorithm}

The odd parities are completely analogous, with their Mellin system:
\begin{equation}
  \setlength{\arraycolsep}{5pt}
  \setlength{\delimitershortfall}{-3pt}
  \frac{\mathrm{d}\boldsymbol{v}}{\mathrm{d}u} = 
  -\begin{bmatrix}
    \dfrac{2u+2x+E}{4u^2+4xu+1} & \dfrac{-\mu}{4u^2+4xu+1} \\[3ex]
    \dfrac{\mu}{4u^2-4xu+1} & \dfrac{2u-2x-E}{4u^2-4xu+1} 
  \end{bmatrix}\boldsymbol{v},
\end{equation}
and with $\chi\in\frac12\mathbb{Z}$ for odd Emary-Bishop states.

\begin{figure}[ht]
  \begin{center}
    \includegraphics[width=.9\textwidth]{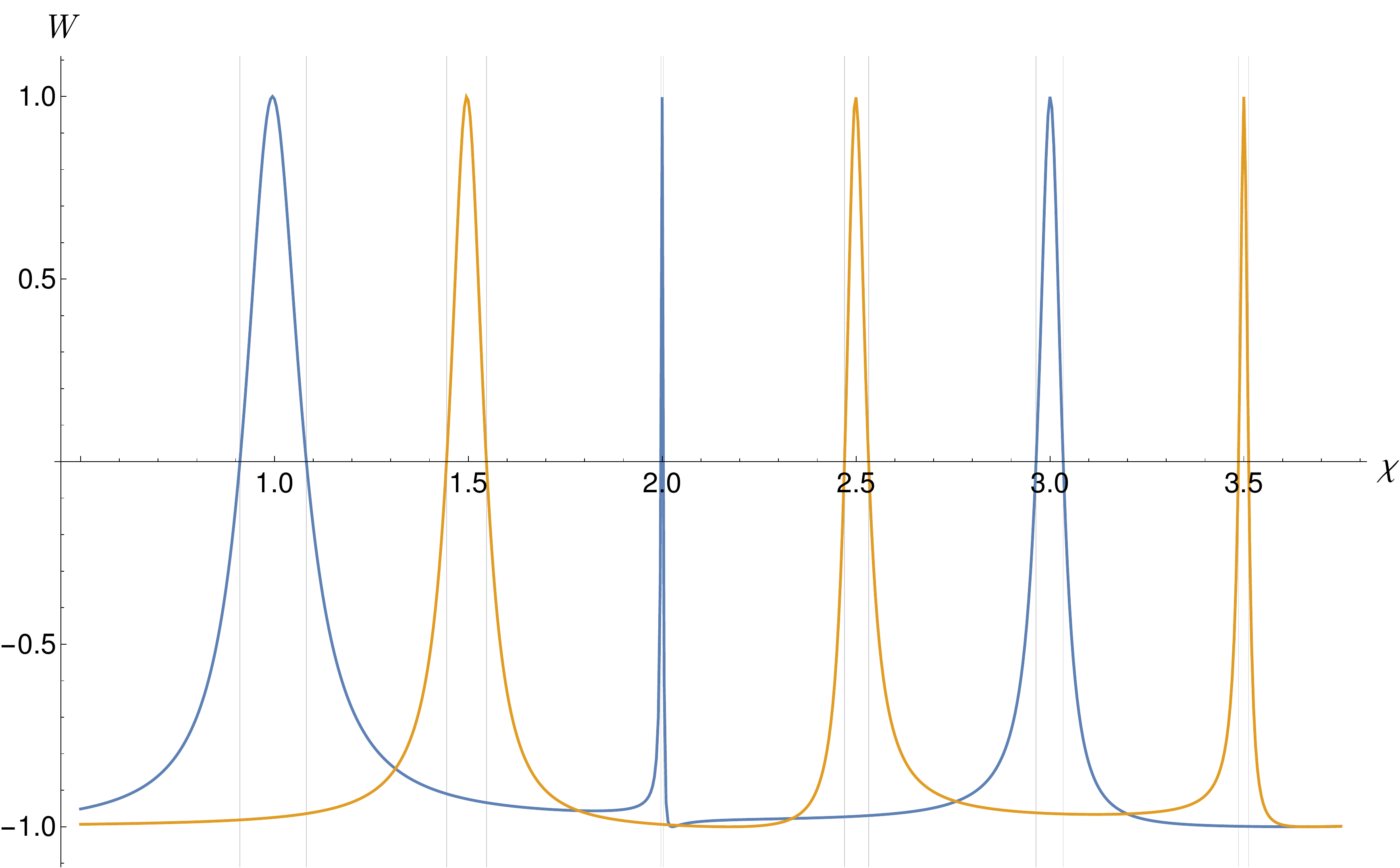}
    \caption{The spectral determinant $W$ as a function of $\chi$ for
      $\kappa=1/2$ and $\mu = 1/3$}
  \label{fig3}
  \end{center}
\end{figure}

\begin{figure}[ht]
  \begin{center}
    \includegraphics[width=.9\textwidth]{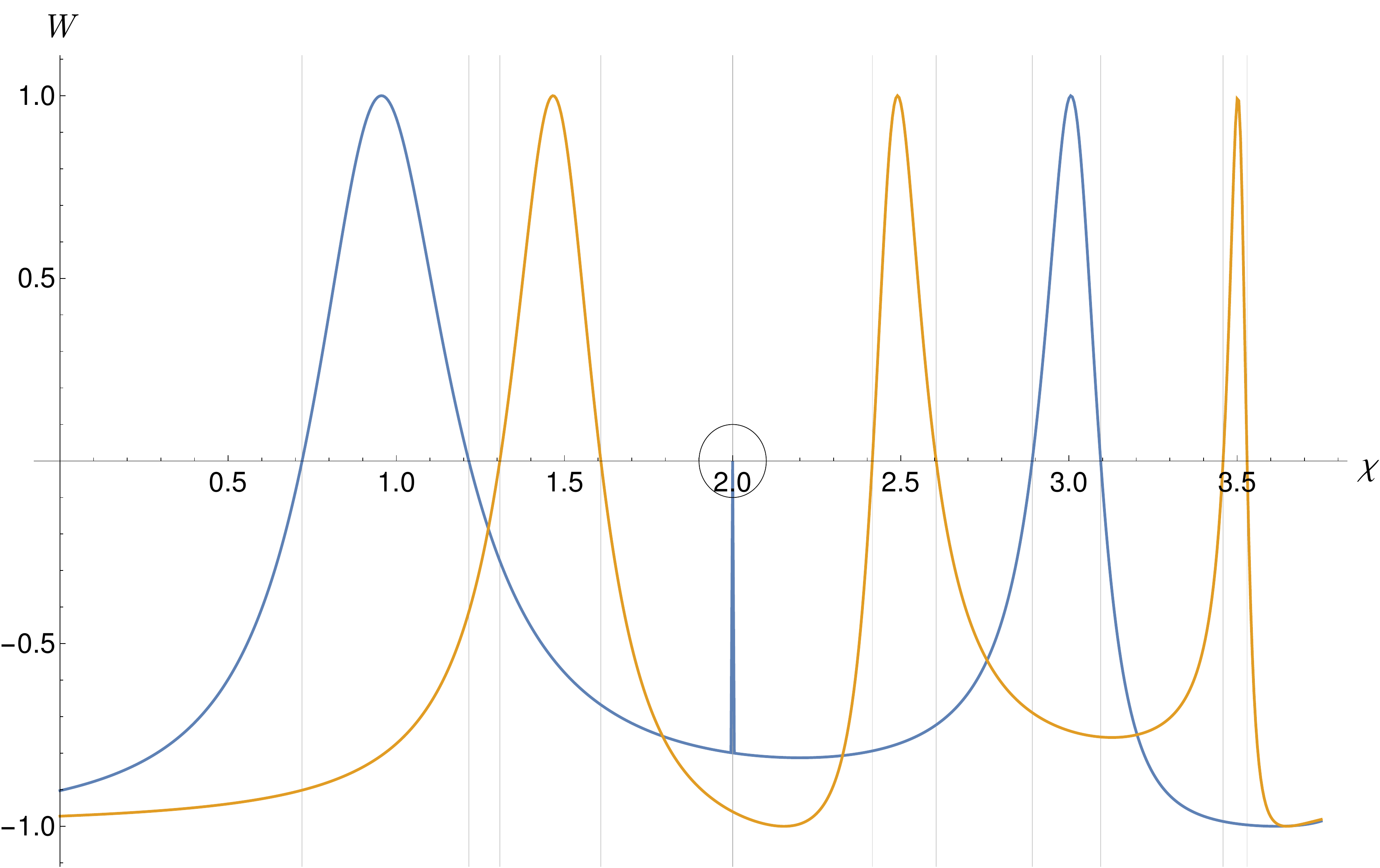}
    \caption{The spectral determinant $W$ as a function of $\chi$ for
      $\kappa=1/2$ and $\mu = 1$. A degenrate Emary-Bishop state is
      present.}
  \label{fig4}
  \end{center}
\end{figure}

A numerical example for a generic situation is presented in Figure
\ref{fig3} and a spectrum with Emary-Bishop states is presented in
Figure \ref{fig4}. We notice in particular, that the function is smooth (or has a removable discontinuity in the degenerate case) which is not the case in other methods which introduce artificial singularities at integer values of the exponent.

\section{Conclusions}

The two photon Rabi model, as formulated in the Bargmann
representation, is unusual in that the respective differential
equation has only entire solutions. The condition that a function is
an eigenstate is reduced only to the finiteness of its norm or, in
other words, the proper asymptotic behaviour at infinity, as specified
by the growth order and type. Whereas in the standard Rabi model one
has to ensure analyticity by gluing together solutions around
different regular singular points, here the problem lies in gluing
solutions with appropriate asymptotic growth.

As infinity is an irregular singular point, in theory such connection
problem would require dealing with the Stokes phenomenon between
formal solutions across the sectors at infinity. However, by using the
Mellin transformation we have shown how to obtain solutions with
prescribed global asymptotics. The intermediate step is the
construction of entire power series, while the transformation is
necessary to select appropriate solutions of the recurrence relation
satisfied by the coefficient of such series.

We note that the starting point of this approach is just the
requirement that the eigen-state be an element of the Bargmann-Fock
space. As opposed to other ad hoc methods in the literature, we thus
arrive at a practical method which is well founded.

The crucial element in the asymptotic analysis are the factorial
series, which, unlike the standard asymptotic expansions, are
convergent. They can be used both for functions of a complex variable
and for solutions of recurrence relations, and they give a concise way
to solve the connection problem or to determine the Stokes phenomenon
as shown in \cite{Okubo:63::}.

Because the system is, in general, not solvable, there are no explicit
elementary formulae for the Stokes multipliers or the connection
coefficients. Thus, even using the factorial series means that
eventually some numerical approximation has to be used. By noticing
that this can be implemented already at the stage of the Mellin
transformation, we further refine our results by investigating how to
give the spectral conditions in terms of contour integrals. These can
then be treated numerically much easier than the relevant infinite
series.

It turns out that the existence of an eigen-state is directly
connected with the properties of the holonomy group of a second order
system of linear differential equations. Furthermore, the
$\mathbb{Z}_4$ symmetry further simplifies the problem, because it
provides a partial connection formula between the holonomy matrices.

Finally, despite the formal development, the holonomy for a linear
system is very easy to compute and leads to a practical Algorithm
\ref{algo}, whose precision is in essence limited only by the
particular chosen scheme of numerical integration.

\acknowledgements
This work has been supported by the grant No. DEC-2011/02/A/ST1/00208 of National Science Centre of Poland.

\appendix

\section{Odd entire solutions}
\label{odd_app}

By defining a new function $f$ such that $z f(z^2) := \psi(z)$ and introducing again $\xi = z^2$ we have the series expansion
\begin{equation}
    f(\xi) = \sum_{n=0}^{\infty}\boldsymbol{c}_n \xi^n,
\end{equation}
and the coefficients satisfy the matrix difference equation
\begin{equation}
2n(2n-1)\boldsymbol{c}_n = 
\begin{bmatrix} E -2x(2n-1) & -\mu \\
\mu & -E +2x(2n-1) \end{bmatrix} \boldsymbol{c}_{n_1} -\boldsymbol{c}_{n-2}.
\end{equation}

Through the same Mellin integral as for the even case we obtain the differential system
\begin{equation}
\frac{\mathrm{d}\boldsymbol{v}}{\mathrm{d}u} =  M (u) \boldsymbol{v}, \quad
  \setlength{\arraycolsep}{5pt}
  \setlength{\delimitershortfall}{-3pt}
    M(u) := -\begin{bmatrix}
    \dfrac{2(u+x)+E}{4u^2+4xu+1} &\dfrac{-\mu}{4u^2+4xu+1}\\[3ex]
    \dfrac{\mu}{4u^2-4xu+1} & \dfrac{2(u-x)-E}{4u^2-xu+1}
    \end{bmatrix},
\end{equation}
whose characteristic exponents are
\begin{equation}
  \begin{aligned}
    \left\{0,\frac12 -\chi\right\}, & \quad \text{for}\quad  u_0=\pm\frac{\kappa}{2}\\
    \left\{0,\chi-1\right\}, &\quad \text{for}\quad    u_0=\pm\frac{1}{2\kappa}:\\
    \left\{-\frac12,-\frac12\right\}, &\quad \text{for}\quad u_0 =
    \infty,
  \end{aligned}
\end{equation}
and the spectral parameter $\chi$ is the same as before.

We note that the logarithmic and Juddian solutions can now arise only for half integer values of $\chi$, and this is the main difference between the even and odd cases.

\section{The case $|\sigma|=\tfrac12$}\label{sigma12}

When $x=1$, there are only two available exponential factors in the
asymptotic expansion: $\exp(\pm\tfrac12 z^2)$, and the convergence of
the Bargmann norm has to be checked in each sector separately. E.g.,
the integral of $\exp(\tfrac12 z^2)$ is finite over the region
$-\pi/4 \leq \arg(z)\leq \pi/4$, but not over
$\pi/4 \leq \arg(z)\leq 3\pi/4$. This means that a normalizable
solution must change its (generalized) type as $\arg(z)$ increases.

If one continues analytically a solution which behaves properly around
the real axis, i.e, $f\sim\exp(-\tfrac12 z^2)$, and there is no Stokes
phenomenon, it will behave as $\exp(\tfrac12|z|^2)$ around the
imaginary axis, and the Bargmann integral will be infinite. A proper
eigenstate cannot have this bahaviour.

To see how the the solution behaves with nontrivial Stokes phenomenon
we can employ the Laplace representation again, which will be
particularly simple for $x=1$. The main equation \eqref{n=2} for
$z^2=\zeta$, which amounts to taking parities $\pm1$ (the
$\pm\mathrm{i}$ case is analogous), is

\begin{equation}
  16\zeta^2 f^{(\mathrm{iv})}+48\zeta f'''+4(3+4\zeta-2\zeta^2)f''+(8-12\zeta)f'
  +(1+\mu^2-4\zeta+\zeta^2)f = 0.
\end{equation}

The integral representation of
\begin{equation}
  f(\zeta) = \int \exp\left(\frac{\zeta}{2}u\right)g(u)\mathrm{d}u,
\end{equation}
gives the following differential equation for $g$
\begin{equation}
  4(1-u^2)^2 g''+ 4(5u-2)(u^2-1)g'+(\mu^2-3-12u+15u^2)= 0,
\end{equation}
whose general solution, for $\mu\neq 1$ is
\begin{equation}
  g = c_1 (1-u)^{\rho}(1+u)^{-\rho-3/2} +
  c_2 (1+u)^{\rho-1}(1-u)^{-\rho-1/2},\quad \rho = -\frac{1+\sqrt{1-\mu^2}}{4},
\end{equation}
or, for $\mu=1$,
\begin{equation}
  g = (1-u)^{-1/4}(1+u)^{-5/4}\left( c_1 + c_2 \log\left(\frac{1+u}{1-u}\right)\right).
\end{equation}

Using the methods of \cite{Maciejewski:15::}, we obtain the positions
of the Stokes lines to be $\arg(z)\in\{\pi/4,-\pi/4,3\pi/4,-3\pi/4\}$,
and that the Stokes matrices are triangular. This means that even if
one chooses a solution with finite partial norm in some sector
\mbox{$S=\{z:\alpha\leq z\leq\beta\}$}
\begin{equation}
  \|f(z)\|_{\alpha,\beta} = \frac{1}{\pi}\int\limits_{S}
  \mathrm{e}^{-|z|^2}|f|\mathrm{d}(\Re(z))\mathrm{d}(\Im(z)),
\end{equation}
its continuation will contain both asymptotics in the next sector
rendering the global integral infinite.

In the generic case, when $x<1$ ($\kappa$ is no longer real), there
are again four exponential types \eqref{types}, except this time they
all lie on the circle $|\sigma|=\tfrac12$ and form a rectangle whose
sides are parallel to the real and imaginary axes. Because the type
alone will not be enough to check the norm, let us go back to equation
\eqref{n=2} and write the solution in the even case (odd being
completely analogous again)
\begin{equation}
  \psi_1(z) = \sum_{n=0}^{\infty} c_n z^{2n},
\end{equation}
whose coefficients correspond to $a_n^1$ of \eqref{fc}. The asymptotic
form of these coefficients can be ascertained either by direct
substitution into the recurrence relation or from the Mellin
representation \eqref{fucktorial}. This time a formal expression is
all we need, because the Bargmann norm will be finite if
\begin{equation}
  \|\psi_1\|=\sum_{n=0}^{\infty}(2n)!|c_n|^2 < \infty,
\end{equation}
and only the behaviour of $c_n$ at infinity matters. Namely, we will
use the Gauss test which states that when
\begin{equation}
  \left|\frac{u_n}{u_{n+1}}\right| = 1 + \frac{h}{n} + \mathscr{O}(n^{-r}),\quad r > 1,
\end{equation}
then the positive series given by $u_n$ converges if and only if
$h>1$.

Since the coefficients in question behave as
\begin{equation}
  c_n \sim \frac{1}{n!}\sigma^n n^\beta,
\end{equation}
its absolute value behaves as
\begin{equation}
  |c_n| = \frac{1}{n!2^n}n^{\Re(\beta)}\left(1+\mathscr{O}(n^{-1})\right),
\end{equation}
where
\begin{equation}
  \beta\in\left\{-\frac14 \pm \frac{E+x}{4\sqrt{x^2-1}},-\frac54 \pm \frac{E+x}{4\sqrt{x^2-1}}\right\},
\end{equation}
in accordance with
\begin{equation}
  \sigma\in\left\{\frac12(-x\pm\sqrt{x^2-1}),\frac12(x\pm\sqrt{x^2-1})\right\}.
\end{equation}
The norm series to be analysed is given by $u_n = (2n)!|c_n|^2$ so
\begin{equation}
  \left|\frac{u_n}{u_{n+1}}\right| = 1 + \frac{1-4\Re(\beta)}{2n}+\mathscr{O}(n^{-2}),
\end{equation}
and the deciding term is $1/n$ for the first two choices of $\beta$
and $3/n$ for the other two.

At this point we have to employ the residual $\mathbb{Z}_2$ symmetry
of the even solutions, because there are more solutions of the
recurrence than of the differential equation. This happens because a
series solution of the differential equaion has imposed on it the
additional conditions $c_n\equiv 0$ for $n<0$. Specifically, we can
only obtain two entire even functions, and there are four pairs of
$(\sigma,\beta)$ specifying asymptotic solutions of the recurrence.

Fortunately it is not necessary to solve the full connection problem,
i.e., decide which asymptotic expansion corresponds to which entire
series. Instead, we recall that if a solution exists, it can be
projected onto parity eigenstates, which satisfy
\begin{equation}
  \psi_1''(z) +2x z\psi_1'(z)+(z^2-E)\psi_1(z)+\frac{\mu}{s}\psi_1(\mathrm{i}z) = 0,\quad s=\pm 1.
\end{equation}
By direct substitution we find that the series coefficients of a
solution of parity $s$ must be a combination of two solutions $c_n$
corresponding to $\sigma$ and $-\sigma$:
\begin{equation}
  d_n = \frac{1}{n!}\sigma^n n^\beta\left(1+\mathscr{O}(n^{-1})\right) 
  - \frac{s\mu}{8x}\frac{1}{n!}(-\sigma)^n n^{\beta-1}\left(1+\mathscr{O}(n^{-1})\right),
\end{equation}
where now there are only two possibilities
\begin{equation}
  \sigma = \frac12(-x\pm\sqrt{x^2-1}),\quad \beta = -\frac14 \pm \frac{E+x}{4\sqrt{x^2-1}},
\end{equation}
so that in effect the asymptotics of $d_n$ is dominated by the larger
$\beta$ and
\begin{equation}
  \left|\frac{u_n}{u_{n+1}}\right| = 1 + \frac{1}{n}+\mathscr{O}(n^{-2}),
\end{equation}
so by Gauss's criterion the norm series is always divergent, proving
no proper eigenstate exists in this case.

\section{Spectral conditions through  factorial series}
\label{app_fac}

The reason why the formula \eqref{fucktorial} can be readily put into practice is that the regular point used, $u_0$, determines
the crucial asymptotic behaviour of the sequence $\boldsymbol{b}_n$.
Because the $\Gamma$ function factors in the sum are of the order
$\mathscr{O}(n^{-1-j-\nu})$, asymptotically one has
\begin{equation}
  \boldsymbol{b}_n \sim \frac{1}{\sqrt{2\pi}} \left(\frac{u_0 \mathrm{e}}{n}\right)^n
  n^{-\nu-3/2}\left(u_0\boldsymbol{h}_0+\mathscr{O}(n^{-1})\right),
\end{equation}
so, by \eqref{ordtype}, $u_0$ is the type of the associated entire
function $f(\xi)$ and also of $\psi(z)$. It will thus suffice to
consider only solutions and contours around the two points
$u_0=\pm\kappa/2$, which give the normalizable types. The other
singular points influence the radius of convergence of \eqref{serg},
so for this series to be integrated term by term over the contour $C$,
the point $u_0$ must lie closer to zero than to any other singular
point, giving the condition $\kappa\leq1/\sqrt{2}$. When this
condition does not hold, a change of variable is required, which
amounts to using a different series for $\boldsymbol{v}$, as explained
in detail in appendix \ref{was_tra}, but in the end $\boldsymbol{b}_n$
is still represented by a factorial series.

When one exponent is $-m\in\mathbb{Z_-}$, the logarithmic solution
$\boldsymbol{v}=\boldsymbol{v}_1 + \log(u-u_0)\boldsymbol{v}_2$ has to
be used, and we notice that continuation around the contour $C$ acts
on this solution as
$\boldsymbol{v}\rightarrow\boldsymbol{v}+2\pi\mathrm{i}\boldsymbol{v}_2$. Both
$\boldsymbol{v}_i$ are single-valued so the contour can be decomposed
into two line segments and an arbitrarily small circle giving
\begin{equation}
  \int_C u^n\boldsymbol{v}\mathrm{d}u =
  \int\limits_0^{u_0-\varepsilon}u^n\log(u-u_0)\boldsymbol{v}_2 \mathrm{d}u +
  \oint\limits_{|u-u_0|=\varepsilon}u^n\boldsymbol{v}\mathrm{d}u +
  \int\limits_{u_0-\varepsilon}^0u^n(\log(u-u_0)+2\pi\mathrm{i})\boldsymbol{v}_2 \mathrm{d}
  =I(\varepsilon),
\end{equation}
and because the exponent of $\boldsymbol{v}_2$ is zero, we can take
the limit
\begin{equation}
  I(\varepsilon)\xrightarrow[\varepsilon\rightarrow 0]{} 2\pi\mathrm{i}\;\text{res}_{u_0}(u^n\boldsymbol{v}_1)
  -2\pi\mathrm{i}\int\limits_0^{u_0}u^n\boldsymbol{v}_2\mathrm{d}u
  =2\pi\mathrm{i}\left(u_0^n \sum_{j=0}^{m-1}\frac{\boldsymbol{h}_{-j-1}n!}{j!(n-j)!}
    -\int\limits_0^{u_0}u^n\boldsymbol{v}_2\mathrm{d}u\right),
\end{equation}
so the situation is the same as before, because the summand behaves as
$\mathscr{O}\left(n^j\right)$, and the integral gives another
factorial series as in \eqref{fucktorial}.

Finally, we remark that the exceptional case when $-m\in\mathbb{Z}$
and the logarithmic term vanishes, corresponds to the Juddian
solutions discovered by Emary and Bishop. This can be verified by
comparing the values of energy numbered by the integer $m$ and the
algebraic conditions on the other parameters which guarantee the
absence of logarithms. The solutions are then of the form
$\exp(\sigma z^2)P(z)$, for a polynomial $P$, so their expansions are
still infinite series, but their Laplace transforms, hence the
solutions $\boldsymbol{v}$, are rational.

Let now $\boldsymbol{b}_n^{\pm}$ denote the solutions of the
recurrence equation \eqref{rec}, constructed by means of the Mellin
transform around $u_0=\pm\kappa/2$, respectively. Their asymptotic
growth is as required, and it remains to be checked whether the
solution $\boldsymbol{a}_n$, obtained around $\xi=0$, is their linear
combination. Because we are dealing with a linear recurrence it is
enough to check the linear dependence for two consecutive elements,
which means that if, for some $n_0\geq2$, the rank of the 4$\times$3
matrix
\begin{equation}
  \begin{bmatrix}
    \boldsymbol{a}_{n_0} & \boldsymbol{b}_{n_0}^+ & \boldsymbol{b}_{n_0}^- \\
    \boldsymbol{a}_{n_0+1} & \boldsymbol{b}_{n_0+1}^+ & \boldsymbol{b}_{n_0+1}^- \\
  \end{bmatrix},
\end{equation}
is less than 3, then
$\boldsymbol{a}_n\in\text{Span}(\boldsymbol{b}_n^+,\boldsymbol{b}_n^-)$. In
practice one has to check that all the 3$\times$3 minors of the above
matrix vanish. If that happens for some value of energy, there exists
an entire solution of the desired asymptotics, i.e, with finite norm.

\section{Factorial series for general position of singular points}\label{was_tra}

Let us deal with the radius of convergence of the series \eqref{serg}. The
crucial obstacle is that both the regular points $\tfrac{\kappa}{2}$ and
$\tfrac{1}{2\kappa}$ can be arbitrarily close to $\tfrac12$ as $\kappa$ gets
close to 1 (and likewise for their negative counterparts), so the radius of
convergence gets smaller and smaller. To remedy
this one can choose the following new independent variable
\begin{equation}
    w = \left(\frac{u}{u_0}\right)^p,
\end{equation}
with a sufficiently large, real $p$. For clarity let us look at the positive
regular point $u_0=\kappa/2$ as the negative case is analogous. The point of
larger absolute value will be mapped into $(1/\kappa^2)^p$, which can be made
larger than 2 by taking 
\begin{equation}
    p>\log_{\tfrac{1}{\kappa^2}}(2) = -\frac{\ln 2}{2\ln\kappa},
\end{equation}
or, for computational purposes,
\begin{equation}
    p = \max\left\{1, \frac{1}{2(1-\kappa)}\right\},
\end{equation}
becasue for $\kappa<2^{-1/2}\approx 0.7$ there radius of convergence is already
large enough, and otherwise we have $-\ln\kappa >1-\kappa$.

\begin{figure}[ht]
\includegraphics[width=\textwidth]{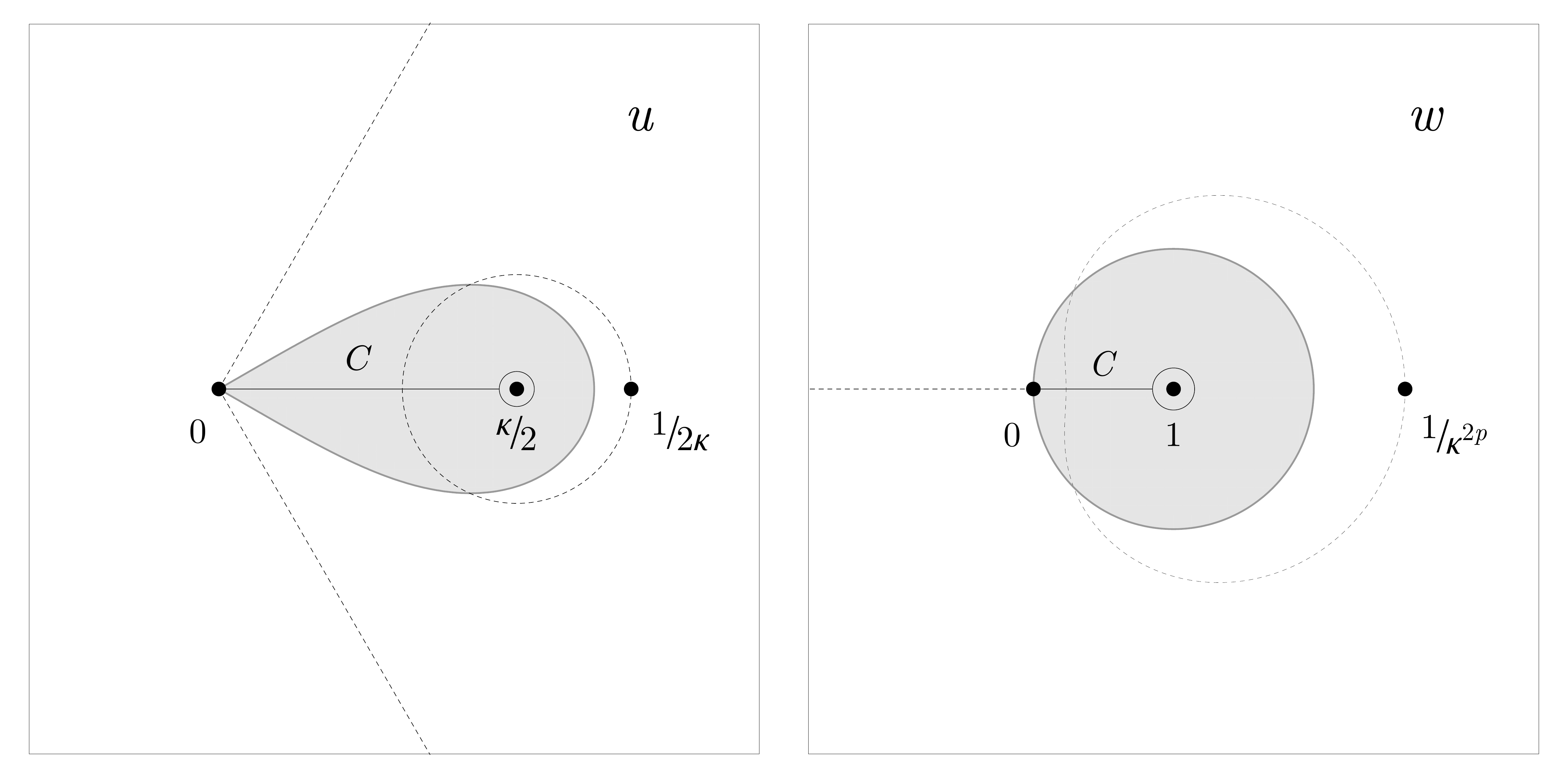}
\caption{The mapping of regions between $u$ and $w$. Dashed circle shows the
disk of convergence of $g(u)$ as the series \eqref{serg}, and the grey region is
where the expansion \eqref{ser2} of $\tilde{g}(w)$ converges. Solid black line
is the integration path $C$ of the Mellin transform.}
\label{regions}
\end{figure}

With proper $p$, the above mapping will send a small connected region around
$u=u_0$ into the disk of radius 1 centered at $w=1$, so the Mellin integral will
change to
\begin{equation}
    \mathcal{M}[v]_n = \frac{u_0^n}{p\,\Gamma(n+1)}\int_C
    w^{\tfrac{n}{p}-1} \tilde v(w)\mathrm{d}w,
\end{equation}
with $\tilde{v}(w) = v(u_0 w^{1/p}) = v(u)$ being holomorphic around $w=1$. An
example of such a disk map is shown in Figure~\ref{regions}. One can then expand
$\tilde{v}(w)$, and tie it with the expansion of $v(u)$
\begin{equation}
    \begin{split}
        \tilde{v}(w) &= \sum_{j=0}^{\infty} H_j (w-1)^{\nu+j} = \\
        v(u_0 w^{1/p} ) &= 
        \sum_{k=0}^{\infty} h_k u_0^{\nu+k} (w^{1/p}-1)^{\nu+k} =
        \sum_{k=0}^{\infty}h_ku_0^{\nu+k} 
        \left( \sum_{l=1}^{\infty}\binom{1/p}{l}(w-1)^l\right)^{\nu+k}.
    \end{split}
    \label{ser2}
\end{equation}
Comparing the two series the following relation between their coefficients can
be found
\begin{equation}
    H_j = u_0^{\nu}\sum_{k=0}^j B_{j-k}\sum_{m=0}^{k} A_{m,k} h_m u_0^{m},
\end{equation}
where $A_{m,j}$ are given recursively by
\begin{equation}
    \begin{aligned}
    A_{0,j} &= \delta_{0j},\\
    A_{1,j} &= \binom{1/p}{j},\\
    A_{m,j} &= \sum_{l=1}^{j-m+1}\binom{1/p}{l}A_{m-1,j-l},
    & \text{for } j\geq m,\\
    A_{m,j} &=0, & \text{for } j<m,
    \end{aligned}
\end{equation}
and $B_j$ are the series coefficient in
\begin{equation}
    (w^{1/p}-1)^{\nu} =: (w-1)^{\nu}\sum_{j=0}^{\infty}B_j(w-1)^j.
\end{equation}

Finally, the modified Mellin transform can be given as
\begin{equation}
    \mathcal{M}[g]_n = \frac{u_0^n}{p\, \Gamma(n+1)} 
    \sum_{j=0}^{\infty}\frac{(-1)^{\nu+j}\Gamma\left(\tfrac{n}{p}+1\right)\Gamma(1+j+\nu)}
    {\Gamma\left(2+j+\tfrac{n}{p}+\nu\right)}H_j.
    \label{fac_ser}
\end{equation}

Alternatively, $H_j$ can be obtained directly, without the use of $h_j$, by writing the system in the variable $w$. Such differential equation has coefficients which are not rational but they admit power series expansion in the relevant region so the solution can be constructed by the Frobenius method around $w=-1$. 


\end{document}